\documentclass[pre,showkeys,amsmath,amssymb]{revtex4}

\usepackage[dvips]{graphicx}
\usepackage{bm}

\newcommand{\beq}[1][\theequation]{\begin{equation}\label{eq:#1}}
\newcommand{\eeq}{\end{equation}}

\newcommand{\nbf}{\mathbf{n}}
\newcommand{\rbf}{\mathbf{r}}
\newcommand{\vbf}{\mathbf{v}}
\newcommand{\ubf}{\mathbf{u}}
\newcommand{\nablabf}{\boldsymbol{\nabla}}
\newcommand{\Gammabf}{\boldsymbol{\Gamma}}
\newcommand{\gammabf}{\boldsymbol{\gamma}}
\newcommand{\Lambdabf}{\boldsymbol{\Lambda}}
\newcommand{\sigmabf}{\boldsymbol{\sigma}}
\newcommand{\Kbf}{\mathbf{K}}
\newcommand{\Lbf}{\mathbf{L}}
\newcommand{\Mbf}{\mathbf{M}}
\newcommand{\Nbf}{\mathbf{N}}
\newcommand{\Ubf}{\mathbf{U}}
\newcommand{\Obf}{\mathbf{0}}

\newcommand{\intO}{\int_{\Omega}\!}
\newcommand{\intdO}{\int_{\partial\Omega}\!\!}
\newcommand{\Matlab}{\textsc{Matlab}}
\newcommand{\Femlab}{\textsc{Femlab}}
\newcommand{\id}{\mathrm{d}}
\newcommand{\half}{\mbox{$\frac{1}{2}$}}

\begin{document}

\title{A high-level programming-language implementation of topology 
optimization applied to steady-state Navier--Stokes flow}

\author{Laurits H{\o}jgaard Olesen, Fridolin Okkels, and 
Henrik Bruus}

\affiliation{MIC -- Department of Micro and Nanotechnology,\\
Technical University of Denmark, DK-2800 Kongens Lyngby, Denmark}

\date{26 June 2005}

\begin{abstract}
We present a versatile high-level programming-language implementation of 
nonlinear topology optimization. Our implementation is based on the 
commercial software package \Femlab{}, and it allows a wide range of 
optimization objectives to be dealt with easily. We exemplify our method 
by studies of steady-state Navier--Stokes flow problems, thus extending the 
work by Borrvall and Petersson on topology optimization of fluids in Stokes 
flow [\emph{Int. J. Num. Meth. Fluids} 2003; \textbf{41}:77--107].
We analyze the physical aspects of the solutions and how they are affected 
by different parameters of the optimization algorithm. A complete example 
of our implementation is included as \Femlab{} code in an appendix.
\end{abstract}

\keywords{topology optimization, Navier--Stokes flow, inertial effects, 
Femlab}

\maketitle

\section{INTRODUCTION}
\label{sec:intro} The material distribution method in topology optimization 
was originally developed for stiffness design of mechanical 
structures~\cite{Bendsoe88} but has now been extended to a multitude of 
design problems in structural mechanics as well as to optics and 
acoustics~\cite{Bendsoe03,Eschenauer,Jensen03,Jensen04}.
Recently Borrvall and Petersson introduced the method for fluids in Stokes 
flow~\cite{Borrvall}.
However, it is desirable to extend the method to fluids described in a full 
Navier--Stokes flow; a direction pioneered by the work of Sigmund and 
Gersborg-Hansen~\cite{Sigmund,GersborgHansen03,GersborgHansen05}.

In the present work we present such an extension by introducing a versatile 
high-level programming-language implementation of nonlinear topology 
optimization, based on the commercial software package \Femlab{}.
It has a wider range of applicability than the Navier--Stokes problems 
studied here, and moreover it allows a wide range of optimization objectives 
to be dealt with easily.

Extending the topology optimization method to new physical domains generally 
involves some rethinking of the design problem and some "trial and error" 
to determine suitable design objectives.
It also requires the numerical analysis and implementation of the problem, 
e.g., using the finite element method (FEM). This process is accelerated a 
lot by using a high-level FEM library or package that allows different
physical models to be joined and eases the tasks of geometry setup, mesh 
generation, and postprocessing.
The disadvantage is that high-level packages tend to have rather complex 
data structure, not easily accessible to the user. This can complicate the 
actual implementation of the problem because the sensitivity analysis is 
traditionally formulated in a low-level manner.

In this work we have used the commercial finite-element package \Femlab{} 
both for the solution of the flow problem and for the sensitivity analysis 
required by the optimization algorithm. We show how this sensitivity 
analysis can be performed in a simple way that is almost independent of 
the particular physical problem studied.
This approach proves even more useful for multi-field extensions, where 
the flow problem is coupled to, e.g., heat conduction, convection-diffusion 
of solutes, and deformation of elastic channel walls in valves and flow 
rectifiers~\cite{Okkels05}.

The paper is organized as follows: In Sec.~\ref{sec:topopt} we introduce 
the topology optimization method for fluids in Navier--Stokes flow, and 
discuss the objective of designing fluidic devices or channel networks for 
which the power dissipation is minimized.
In Sec.~\ref{sec:general} we express the Navier--Stokes equations in a 
generic divergence form that allows them to be solved with \Femlab{}. 
This form encompasses a wide range of physical problems. We also work out 
the sensitivity analysis for a class of integral-type optimization 
objectives in such a way that the built-in symbolic differentiation tools 
of \Femlab{} can be exploited.
In Sec.~\ref{sec:examples} we present our two numerical examples that
illustrates different aspects and problems to consider: The first example 
deals with designing a structure that can guide the flow in the reverse 
direction of an applied pressure drop. The general outcome of the 
optimization is an $S$-shaped channel, but the example illustrates how the 
detailed structure depends on the choice of the parameters of the algorithm. 
The second example deals with a four terminal device where the fluidic 
channel design that minimizes the power dissipation shows a Reynolds number 
dependence. As the Reynolds number is increased a transition occurs between 
two topologically different solutions, and we discuss how the position of 
the transition depends on the choice of initial conditions.
Finally in the appendix we include a transcript of our \Femlab{} code 
required for solving the second numerical example. The code amounts to 111 
lines -- excluding the optimization algorithm that can be obtained by 
contacting K.~Svanberg~\cite{Svanberg88,Svanberg02,mailto:Svanberg}.

\section{TOPOLOGY OPTIMIZATION FOR NAVIER--STOKES FLOW IN STEADY STATE}
\label{sec:topopt} Although our high-level programming-language 
implementation is generally applicable we have chosen to start on the 
concrete level by treating the basic equations for our main example: the 
full steady-state Navier--Stokes flow problem for incompressible fluids.

We consider a given computational domain $\Omega$ with appropriate boundary 
conditions for the flow given on the domain boundary $\partial\Omega$.
The goal of the optimization is to distribute a certain amount of solid
material inside $\Omega$ such that the material layout defines a fluidic
device or channel network that is optimal with respect to some objective, 
formulated as a function of the variables, e.g., minimization of the power 
dissipated inside the domain.

The basic principle in the material distribution method for topology
optimization is to replace the original discrete design problem with a 
continuous one where the material density is allowed to vary continuously 
between solid and void~\cite{Bendsoe03}.
Thus in our flow problem we assume the design domain to be filled with some 
idealized porous material of spatially varying permeability. Solid wall and 
open channels then correspond to the limits of very low and very high 
permeability, respectively.

In the final design there should preferably be no regions at intermediate 
permeability since otherwise it cannot be interpreted as a solution to the 
original discrete problem. Alternatively it may be possible to fabricate the 
device from polymeric materials such as PDMS that naturally have a finite 
permeability to the fluid~\cite{Geschke03}.

\subsection{Governing equations for flow in idealized porous media}
\label{sec:GovEqs} We assume that the fluid flowing in the idealized 
porous medium is subject to a friction force~$\mathbf{f}$ which is 
proportional to the fluid velocity $\vbf$, c.f. Darcy's law.
Thus $\mathbf{f}=-\alpha\vbf$, where $\alpha(\rbf)$ is the inverse of the 
local permeability of the medium at position $\rbf$. These properties of 
the idealized porous medium may only be approximately valid for an actual
medium. However, the assumptions are not in conflict with any fundamental 
physical law, and since the converged solutions contain only solid walls 
and open channels, the specific nature of the idealized porous medium is 
of no consequence.

The flow problem is described in terms of the fluid velocity field 
$\vbf(\rbf)$ and pressure $p(\rbf)$. The governing equations are the 
steady state Navier--Stokes equation and the incompressibility constraint
 \begin{alignat}{2}
 \rho(\vbf\cdot\nablabf)\vbf &= \nablabf\cdot\sigmabf - \alpha\vbf,
  \label{eq:NavierStokes} \\
 \nablabf\cdot\vbf &= 0, \label{eq:div}
 \end{alignat}
where $\rho$ is the mass density of the fluid. For an incompressible
Newtonian fluid the components $\sigma_{ij}$ of the Cauchy stress tensor
$\sigmabf$ are given by
 \beq[cauchy]
 \sigma_{ij} = -p\: \delta_{ij} + \eta\Big(\frac{\partial v_i}{\partial x_j}
  + \frac{\partial v_j}{\partial x_i}\Big),
 \eeq
where $\eta$ is the dynamic viscosity. The formalism is valid in three 
dimensions, but for simplicity we shall consider only two-dimensional 
problems, i.e., we assume translational invariance in the third dimension 
and set $\rbf = (x_1,x_2)$ and $\vbf=\big(v_1(\rbf),v_2(\rbf)\big)$.
The boundary conditions will typically be either Dirichlet type specifying 
the velocity field $\vbf$ on the boundary or Neumann type specifying the 
external forces $\nbf\cdot\sigmabf$.

It is convenient to introduce a design variable field $\gamma(\rbf)$
controlling the local permeability of the medium. We let $\gamma$ vary 
between zero and unity, with $\gamma=0$ corresponding to solid material 
and $\gamma=1$ to no material.
Following Ref.~\cite{Borrvall} we then relate the local inverse 
permeability $\alpha(\rbf)$ to the design field $\gamma(\rbf)$ by the 
convex interpolation
 \beq[alpha]
 \alpha(\gamma) \equiv \alpha_{\min} + (\alpha_{\max} - \alpha_{\min})\:
  \frac{q\big[1-\gamma\big]}{q+\gamma},
 \eeq
where $q$ is a real and positive parameter used to tune the shape of 
$\alpha(\gamma)$. Ideally, impermeable solid walls would be obtained 
with $\alpha_{\max}=\infty$, but for numerical reasons we need to 
choose a finite value for $\alpha_{\max}$. For the minimal value we 
choose $\alpha_{\min}=0$.%
\footnote{Borrvall and Petersson suggest a model for plane flow between 
two parallel surfaces of varying separation $h(\rbf)$. The power 
dissipation due to out-of-plane shears is modelled by an absorption term 
$-\alpha\vbf$, where $\vbf(\rbf)$ is the average velocity between the 
surfaces and $\alpha(\rbf) = 12\eta/h(\rbf)^2$.
In their model it is therefore natural to operate with a non-zero 
$\alpha_{\min}=12\eta/h_{\max}^2$ in Eq.~\eqref{eq:alpha}.}

For a given material distribution $\gamma(\rbf)$ there are two 
dimensionless numbers characterizing the flow, namely the Reynolds number
 \beq[reynoldnumber]
 Re = \frac{\rho\,\ell\,v}{\eta}
 \eeq
describing the ratio between inertia and viscous forces, and the Darcy 
number
 \beq[darcynumber]
 Da = \frac{\eta}{\alpha_{\max}\ell^2}
 \eeq
describing the ratio between viscous and porous friction forces. 
Here $\ell$ is a characteristic length scale of the system and $v$ a
characteristic velocity.

Almost impermeable solid material is obtained for very low Darcy numbers, 
in practice $Da\lesssim 10^{-5}$. Further insight into the meaning of the 
Darcy number is gained by considering Poiseuille flow in a channel or 
slit of width $\ell$ between two infinite parallel plates of porous 
material. In this case the fluid velocity inside the porous walls decays 
on a length scale $\ell^{{}}_{Da}$, where 
$\ell^{{}}_{Da} = \sqrt{Da}\,\ell = \sqrt{\eta/\alpha_{\max}}$.
See also Sec.~\ref{sec:ReverseFlowStokes} for details on how the flow 
depends on~$Da$.

\subsection{Power dissipation}
In the pioneering work by Borrvall and Petersson~\cite{Borrvall} the main 
focus was on minimizing the power dissipation in the fluid.
The total power $\Phi$ dissipated inside the fluidic system (per unit 
length in the third dimension) is given by~\cite{Landau}
 \begin{subequations}
 \beq[power]
 \Phi(\vbf,p,\gamma)
  = \intO\bigg[\half\eta\sum_{i,j}
     \Big(\frac{\partial v_i}{\partial x_j} 
      + \frac{\partial v_j}{\partial x_i}\Big)^2
  + \sum_i\alpha(\gamma) v_i^2\bigg]\,\id\rbf.
 \eeq
In steady-state this is equal to the sum of the work done on the system 
by the external forces and the kinetic energy convected into it,
 \beq[work]
 \Phi(\vbf,p,\gamma) = 
  \int_{\partial\Omega}\sum_{i,j}\Big[n_i\sigma_{ij}v_j
  - n_iv_i\big(\half\rho v_j^2\big)\Big]\,\id s.
 \eeq
Here $\nbf$ is a unit outward normal vector such that $\nbf\cdot\sigmabf$ 
is the external force acting on the system boundary and 
$\nbf\cdot\sigmabf\cdot\vbf$ is the work done on the system by this force.
Moreover, in the common case where the geometry and boundary conditions
are such that the no-slip condition $\vbf = \Obf$ applies on all external
solid walls, while on the inlet and outlet boundaries $\vbf$ is parallel 
to $\nbf$ and $(\nbf\cdot\nablabf)\,\vbf = 0$,%
\footnote{In particular this is the case when the inlets and outlets are 
chosen as straight channels sufficiently long that prescribing a parabolic 
Poiseuille profile can be justified, see Figs.~\ref{fig:Sgeom} 
and~\ref{fig:setup}.}
Eq.~\eqref{eq:work} reduces to
 \beq[work2]
 \Phi(\vbf,p,\gamma) = 
  \intdO-\nbf\cdot\vbf\,\big(p + \half\rho v^2\big)\,\id s.
 \eeq
 \end{subequations}
Borrvall and Petersson showed that for Stokes flow with Dirichlet 
boundary conditions everywhere on the boundary $\partial\Omega$, the 
problem of minimizing the total power dissipation inside the fluidic 
device subject to a volume constraint on the material distribution is 
mathematically well-posed.
Moreover it was proven that in the case where $\alpha(\gamma)$ is a
linear function, the optimal material distribution is fully 
discrete-valued.

When $\alpha(\gamma)$ is not linear but \emph{convex} then the solid/void 
interfaces in the optimal solution are not discrete zero/unity 
transitions but slightly smeared out. Convexity implies that the (negative 
value of the) slope of $\alpha$ at $\gamma=0$ is larger than at $\gamma=1$; 
therefore there will be a neighbourhood around the discrete interface where 
it pays to move material from the solid side to the void.
Using the interpolation in Eq.~\eqref{eq:alpha} we have
$\alpha'(0)=(\alpha_{\min}-\alpha_{\max})\frac{1+q}{q}$ and
$\alpha'(1)=(\alpha_{\min}-\alpha_{\max})\frac{q}{1+q}$.
For large values of $q$ the interpolation is almost linear and we expect
almost discrete interfaces, whereas for small $q$ we expect smeared out 
interfaces in the optimized solution.

Consider the case when Eq.~\eqref{eq:work2} applies. If the system is 
driven with a prescribed flow rate then minimizing the total power 
dissipation is clearly equivalent to minimizing the pressure drop across 
the system.
Conversely, if the system is driven at a prescribed pressure drop, then 
the natural design objective will be to maximize the flow rate which is
equivalent to maximizing the dissipated power, c.f. Eq.~\eqref{eq:work2}.
In either case the objective can be described as minimizing the
\emph{hydraulic resistance} of the system.

For problems with more complex design objectives, such as a minimax 
problem for the flow rate through several different outlets, there will 
typically be no analog in terms of total dissipated power.
In such cases there is no guarantee for the existence of a unique optimal
solution and one has to be extra careful when formulating the design 
problem.

\section{GENERALIZED FORMULATION OF THE OPTIMIZATION PROBLEM}
\label{sec:general} For a given material distribution we solve the
Navier--Stokes flow problem using the commercial finite element software 
\Femlab{}. It provides both a graphical front-end and a library of 
high-level scripting tools based on the \Matlab{} programming language, 
and it allows the user to solve a wide range of physical problems by 
simply typing in the strong form of the governing equations as text 
expressions. The equations must then comply with a generic divergence 
form that eases the conversion to weak form required for the finite 
element solution. However, that is not a severe constraint since this 
is the natural way of expressing most partial differential equations 
originating from conservation laws.

Since we have chosen fluidics as our main example, we begin by expressing 
the incompressible Navier--Stokes flow problem in divergence form.
Then we state the optimization problem with a general form of the design 
objective function and perform the discretization and sensitivity analysis 
based on this generalized formulation.
We stress that although for clarity our examples are formulated in two 
dimensions only, the method is fully applicable for 3D systems.

\subsection{The flow problem in divergence form}
\label{sec:divergenceform} We first introduce the velocity-pressure vector 
$\ubf = [v_1,\: v_2,\: p\:]$ and define for $i=1,2,3$ the quantities 
$\Gammabf_i$ and $F_i$ as
 \beq
 \Gammabf_1 \equiv
  \begin{bmatrix}\sigma_{11}\\ \sigma_{21}\end{bmatrix}, \quad
 \Gammabf_2 \equiv
  \begin{bmatrix}\sigma_{12}\\ \sigma_{22}\end{bmatrix}, \quad
 \Gammabf_3 \equiv \begin{bmatrix}0\\ 0\end{bmatrix},
 \eeq
and
 \beq
 F_1 \equiv \rho(\vbf\cdot\nablabf)v_1 + \alpha(\gamma) v_1, \quad
 F_2 \equiv \rho(\vbf\cdot\nablabf)v_2 + \alpha(\gamma) v_2, \quad
 F_3 \equiv \nablabf\cdot\vbf.
 \eeq
Using this, Eqs.~\eqref{eq:NavierStokes} and \eqref{eq:div} can be 
written in divergence form as
 \begin{subequations}
 \label{eq:generic}
 \begin{alignat}{3}
 \nablabf\cdot\Gammabf_i &= F_i &\mbox{in }&\Omega,
  &\quad&\textit{ Governing equations} \label{eq:divform}\\
 R_i &= 0 &\mbox{on }& \partial\Omega,
  &&\textit{ Dirichlet b.c.} \rule{0mm}{6mm}\label{eq:dirichlet}\\
 -\nbf\cdot\Gammabf_i &= G_i +
  \smash[t]{\sum_{j=1}^3\frac{\partial R_j}{\partial u_i}\mu_j}
  & \qquad \mbox{on }& \partial\Omega,
  &&\textit{ Neumann b.c.} \label{eq:neumann}
 \end{alignat}
 \end{subequations}
where $\Gammabf_i$ and $F_i$ are understood to be functions of the
solution $\ubf$, its gradient $\nablabf\ubf$, and of the design variable 
$\gamma$. The quantity $R_i(\ubf,\gamma)$ in Eq.~\eqref{eq:dirichlet} 
describes Dirichlet type boundary conditions. For example, fluid no-slip 
boundary conditions are obtained by defining $R_1\equiv v_1$ and 
$R_2\equiv v_2$ on the external solid walls. 
The quantity $G_i(\ubf,\gamma)$ in Eq.~\eqref{eq:neumann} describe 
Neumann type boundary conditions, and $\mu_i$ denote the Lagrange 
multiplier necessary to enforce the constraint $R_i = 0$, e.g., the 
force with which the solid wall has to act upon the fluid to enforce the 
no-slip boundary condition.
Of course, it is not possible to enforce both Dirichlet and Neumann 
boundary conditions for the same variable simultaneously.
Only when the variable $u_i$ is not fixed by any of the Dirichlet
constraints $R_j$ does the Neumann condition $G_i$ come into play,
as all $\partial R_j/\partial u_i$ vanish and the Lagrange multipliers 
$\mu_j$ are decoupled from Eq.~\eqref{eq:neumann}.
Inactive Dirichlet constraints can be obtained simply by specifying
the zero-function $R_i\equiv0$, that also satisfies 
Eq.~\eqref{eq:dirichlet} trivially.

\subsection{The objective function}
In general the design objective for the optimization is stated as the
minimization of a certain \emph{objective function} $\Phi(\ubf,\gamma)$.
We shall consider a generic integral-type objective function of the form
 \beq[ABPhi]
 \Phi(\ubf,\gamma) = \intO A(\ubf,\gamma)\,\id\rbf +
 \intdO B(\ubf,\gamma)\,\id s.
 \eeq
In particular, we can treat the design objective of minimizing the power
dissipation inside the fluidic domain by taking, c.f. Eq.~\eqref{eq:power}
 \beq
 A \equiv
 \half\eta\sum_{i,j}\Big(\frac{\partial v_i}{\partial x_j}
  + \frac{\partial v_j}{\partial x_i}\Big)^2
  + \sum_i\alpha(\gamma) v_i^2 \quad\textrm{in }\Omega\qquad\textrm{and}
 \qquad B\equiv0\quad \textrm{on }\partial\Omega.
 \eeq
Alternatively, the objective of maximizing the flow out through a 
particular boundary segment $\partial\Omega_o$ is obtained by choosing
 \beq
 A \equiv 0 \quad \textrm{in }\Omega \qquad\textrm{and}\qquad
 B \equiv
  \left\{\begin{array}{cl}-\nbf\cdot\vbf &\quad 
   \textrm{on }\partial\Omega_o, \\
   0 &\quad \textrm{on }\partial\Omega\backslash\partial\Omega_o,
  \end{array}\right.
 \eeq
and objectives related to $N$ discrete points $\rbf_k$ can be treated 
using Dirac delta functions as
 \beq
 A\equiv\sum_{k=1}^N A_k(\ubf,\gamma)\: \delta(\rbf-\rbf_k)
 \quad\textrm{in }\Omega \qquad \textrm{and} \qquad
 B\equiv0 \quad\textrm{on }\partial\Omega.
 \eeq
Finally we stress that not all optimization objectives lend themselves 
to be expressed in the form of Eq.~\eqref{eq:ABPhi} -- an example of 
which is the problem of maximizing the lowest vibrational eigenfrequency 
in structural mechanics.

\subsection{Optimization problem}
The optimal design problem can now be stated as a continuous constrained 
nonlinear optimization problem:
 \begin{subequations}
 \begin{alignat}{3}
 \min_{\gamma}\: \Phi(\ubf,\gamma)& \\[-1mm]
 \textrm{subject to}\: 
 :&\ \intO \gamma(\rbf)\id\rbf - \beta|\Omega| \leq 0,
  &\qquad&\textit{Volume constraint} \label{eq:stvol}\\
 :&\quad 0\leq\gamma(\rbf)\leq 1,
  &\qquad&\textit{Design variable bounds}\label{eq:stbounds}\\[2mm]
 :&\quad \textrm{Eqs.~\eqref{eq:divform} to \eqref{eq:neumann}},
  &\qquad&\textit{Governing equations} \label{eq:stgov}
 \end{alignat}
 \end{subequations}
With the volume constraint we require that at least a fraction $1-\beta$ 
of the total volume $|\Omega|$ should be filled with porous material.

The very reason for replacing the original discrete design problem with 
a continuous one by assuming a porous and permeable material, is that it 
allows the use of efficient mathematical programming methods for smooth 
problems. We have chosen the popular method of moving asymptotes 
(MMA)~\cite{Svanberg88,Svanberg02}, which is designed for problems with 
a large number of degrees-of-freedom and thus well-suited for topology 
optimization~\cite{Bendsoe03}.
It is a gradient-based algorithm requiring information about the 
derivative with respect to $\gamma$ of both the objective function
$\Phi$ and the constraints. Notice that for any $\gamma$ the governing 
equations allow us to solve for $\ubf$; therefore in effect they define 
$\ubf[\gamma]$ as an implicit function. The gradient of $\Phi$ is then 
obtained using the chain rule
 \beq[gradPhi]
 \frac{d}{d\gamma}\Big[\Phi\big(\ubf[\gamma],\gamma\big)\Big] =
  \frac{\partial\Phi}{\partial\gamma} +
  \int_\Omega \frac{\partial\Phi}{\partial\ubf}\cdot
   \frac{\partial\ubf}{\partial\gamma}\,\id\rbf.
 \eeq
However, because $\ubf[\gamma]$ is implicit, it is impractical to 
evaluate the derivative $\partial\ubf/\partial\gamma$ directly.
Instead, we use the adjoint method to eliminate it from 
Eq.~\eqref{eq:gradPhi} by computing a set of Lagrange multipliers for
Eqs.~\eqref{eq:divform} to \eqref{eq:neumann} considered as
constraints~\cite{Michaleris}.
For details see Sec.~\ref{sec:discretization}.

The optimization process is iterative and the $k$th iteration consists of
three steps:
\begin{itemize}
\item[(i)] Given a guess $\gamma^{(k)}$ for the optimal material 
distribution we first solve Eqs.~\eqref{eq:divform} to \eqref{eq:neumann}
for $\ubf^{(k)}$ as a finite element problem using \Femlab{}.
\item[(ii)] Next, the sensitivity analysis is performed where the 
gradient of the objective and constraints with respect to $\gamma$ is 
evaluated. In order to eliminate $\partial\ubf/\partial\gamma$ from 
Eq.~\eqref{eq:gradPhi} we solve the adjoint problem of 
Eqs.~\eqref{eq:divform} to \eqref{eq:neumann} for the Lagrange multipliers 
$\widetilde{\ubf}^{(k)}$, also using \Femlab{}.
\item[(iii)] Finally, we use MMA to obtain a new guess $\gamma^{(k+1)}$ 
for the optimal design based on the gradient information and the past 
iteration history.
\end{itemize}
Of the three steps, (i) is the most expensive computationalwise since it 
involves the solution of a nonlinear partial differential equation.

\subsection{Discretization and sensitivity analysis}
\label{sec:discretization} The starting point of the finite element 
analysis is to approximate the solution component $u_i$ on a set of 
finite element basis functions $\{\varphi_{i,n}(\rbf)\}$,
 \beq
 u_i(\rbf) = \sum_n u_{i,n}\,\varphi_{i,n}(\rbf),
 \eeq
where $u_{i,n}$ are the expansion coefficients. Similiarly, the design 
variable field $\gamma(\rbf)$ is expressed as
 \beq
 \gamma(\rbf) = \sum_n\gamma_n\,\varphi_{4,n}(\rbf).
 \eeq
For our incompressible Navier--Stokes problem we use the standard 
Taylor--Hood element pair with quadratic velocity approximation and
linear pressure. For the design variable we have chosen the linear 
Lagrange element.%
\footnote{Another common choice is the discontinuous and piecewise
constant element for the design variable. Notice that for second and 
higher order Lagrange elements the condition $0\leq\gamma_n\leq1$ does 
not imply $0\leq\gamma(\rbf)\leq1$ for all $\rbf$ because of overshoot 
at sharp zero-to-unity transitions in $\gamma$. This in turn can result 
in negative $\alpha$, c.f. Eq.~\eqref{eq:alpha}, which is unphysical and 
also destroys the convergence of the algorithm.}

The problem Eqs.~\eqref{eq:divform} to \eqref{eq:neumann} is discretized
by the Galerkin method and takes the form
 \beq[Li+Mi]
 \Lbf_i(\Ubf,\gammabf) - \sum_{j=1}^3\Nbf_{ji}^T\Lambdabf_j = \Obf
 \qquad \mbox{and} \qquad
 \Mbf_i(\Ubf,\gammabf) = \Obf,
 \eeq
where $\Ubf_i$, $\Lambdabf_i$, and $\gammabf$ are column vectors holding 
the expansion coefficients for the solution~$u_{i,n}$, the Lagrange 
multipliers $\mu_{i,n}$, and the design variable field $\gamma_n$, 
respectively.
The column vector $\Lbf_i$ contains the projection of 
Eq.~\eqref{eq:divform} onto $\varphi_{i,n}$ which upon partial 
integration is given by
 \beq[Lint]
 \Lbf_{i,n} = \intO\big(\varphi_{i,n}\,F_i +
  \nablabf\varphi_{i,n}\cdot\Gammabf_i)\,\id\rbf
  + \intdO \varphi_{i,n}\,G_i\,\id s.
 \eeq
The column vector $\Mbf_i$ contains the pointwise enforcement of the
Dirichlet constraint Eq.~\eqref{eq:dirichlet}
 \beq
 \Mbf_{i,n} = R_i\big(\ubf(\rbf_{i,n})\big).
 \eeq
Finally, the matrix $\Nbf_{ij}=-\partial\Mbf_i/\partial\Ubf_j$ describes 
the coupling to the Lagrange multipliers in Eq.~\eqref{eq:neumann}.
The solution of the nonlinear system in Eq.~\eqref{eq:Li+Mi} above 
corresponds to step (i) in $k$th iteration. The sensitivity analysis in
step (ii) requires us to compute
 \begin{subequations}
 \beq[gradPhi2]
 \frac{d}{d\gammabf}\Big[\Phi\big(\Ubf(\gammabf),\gammabf\big)\Big] =
  \frac{\partial\Phi}{\partial\gammabf} +
  \sum_{i=1}^3\frac{\partial\Phi}{\partial\Ubf_i}
  \frac{\partial\Ubf_i}{\partial\gammabf},
 \eeq
which is done using the standard adjoint method~\cite{Michaleris}.
By construction we have for any $\gammabf$ that 
$\Lbf_i\big(\Ubf(\gammabf),\gammabf\big)
 -\sum_{j=1}^3\Nbf_{ji}^T\Lambdabf_j(\gammabf) = \Obf$
and $\Mbf_i\big(\Ubf(\gammabf),\gammabf\big) = \Obf$.
Therefore also the derivative of those quantities with respect to 
$\gammabf$ is zero, and adding any multiple, say $\widetilde{\Ubf}_i$ 
and $\widetilde{\Lambdabf}_i$, of them to Eq.~\eqref{eq:gradPhi2} does 
not change the result
 \begin{align}
 \frac{d}{d\gammabf}\Big[\Phi\big(\Ubf(\gammabf),\gammabf\big)\Big]
 =& \frac{\partial\Phi}{\partial\gammabf}
  + \sum_{i=1}^3\frac{\partial\Phi}{\partial\Ubf_i}
    \frac{\partial\Ubf_i}{\partial\gammabf}
  + \sum_{i=1}^3\bigg[\widetilde{\Ubf}_i^T\frac{\partial}{\partial\gammabf}
    \Big(\Lbf_i-\sum_{j=1}^3\Nbf_{ji}^T\Lambdabf_j\Big)
  - \widetilde{\Lambdabf}_i^T\frac{\partial}{\partial\gammabf}
    \Big(\Mbf_i\Big)\bigg]\nonumber\\
 =& \frac{\partial\Phi}{\partial\gammabf}
  + \sum_{i=1}^3\Big(\widetilde{\Ubf}_i^T 
   \frac{\partial\Lbf_i}{\partial\gammabf}
  - \widetilde{\Lambdabf}_i^T
   \frac{\partial\Mbf_i}{\partial\gammabf}\Big)\nonumber\\
 &+ \sum_{i=1}^3\bigg[\frac{\partial\Phi}{\partial\Ubf_i}
  + \sum_{j=1}^3\Big(\widetilde{\Ubf}_j^T
   \frac{\partial\Lbf_j}{\partial\Ubf_i} +
   \widetilde{\Lambdabf}_j^T\Nbf_{ji}\Big)\bigg]
   \frac{\partial\Ubf_i}{\partial\gammabf}
  - \sum_{i=1}^3\bigg[\sum_{j=1}^3\widetilde{\Ubf}_j^T\Nbf_{ij}^T\bigg]
   \frac{\partial\Lambdabf_i}{\partial\gammabf}.\label{eq:undesired}
 \end{align}
 \end{subequations}
Here we see that the derivatives $\partial\Ubf_i/\partial\gammabf$ and
$\partial\Lambdabf_i/\partial\gammabf$ of the implicit functions can be
eliminated by choosing $\widetilde{\Ubf}_i$ and $\widetilde{\Lambdabf}_i$
such that
 \beq[adjoint]
 \sum_{j=1}^3\Big(\Kbf_{ji}^T\widetilde{\Ubf}_j
  - \Nbf_{ji}^T\widetilde{\Lambdabf}_j\Big)
  = \frac{\partial\Phi}{\partial\Ubf_i}
 \qquad\mbox{and}\qquad
 \sum_{j=1}^3\Nbf_{ij}\widetilde{\Ubf}_j = \Obf,
 \eeq
where we introduced $\Kbf_{ij}=-\partial\Lbf_i/\partial\Ubf_j$.
This problem is the adjoint of Eq.~\eqref{eq:Li+Mi} and
$\widetilde{\Ubf}$ and $\widetilde{\Lambdabf}$ are the corresponding 
Lagrange multipliers.

In deriving Eq.~\eqref{eq:undesired} we implicitly assumed that 
$\Nbf_{ij}$ is independent of $\gammabf$, i.e., that the constraint 
$R_i(\ubf,\gamma)$ is a linear function.
If this is not true then the gradient $\partial\Phi/\partial\gammabf$ 
computed from Eq.~\eqref{eq:undesired} is not exact, which may leed to 
poor performance of the optimization algorithm if the constraints are 
strongly nonlinear.
In order to avoid such problems it is necessary to include the nonlinear
parts of the constraint vector $\Mbf$ into $\Lbf$ and move the 
corresponding Lagrange multipliers from $\Lambdabf$ into $\Ubf$.
While this is beyond the scope of the divergence form discussed in 
Sec.~\ref{sec:divergenceform}, it is certainly possible to deal with 
such problems in \Femlab{}. Also the sensitivity analysis above remains 
valid since it relies only on the basic form of Eq.~\eqref{eq:Li+Mi} for 
the discretized problem.

\subsection{Implementation aspects}
We end this section by discussing a few issues on the implementation of
topology optimization using \Femlab{}.

Firstly there is the question of how to represent the design variable 
$\gamma(\rbf)$. The governing equations as expressed by $\Gammabf_i$ and 
$F_i$ in Eq.~\eqref{eq:divform} depend not only on the solution $\ubf$ but 
also on $\gamma$, and the implementation should allow for this dependence 
in an efficient way. Here our simple and straightforward approach is to 
include $\gamma$ as an extra dependent variable on equal footing with the
velocity field and pressure, i.e., we append it to the velocity-pressure 
vector, redefining $\ubf$ as
 \beq
 \ubf\equiv[v_1,v_2,p,\gamma].
 \eeq
This was already anticipated when we denoted the basis set for $\gamma$ 
by $\{\phi_{4,n}(\rbf)\}$. By making $\gamma$ available as a field
variable we can take full advantage of all the symbolic differentiation, 
matrix, and postprocessing tools for analysing and displaying the material 
distribution. Appending $\gamma$ to the list of dependent variables we are 
required to define a fourth governing equation. However, since we are 
never actually going to solve this equation, but rather update $\gamma$ 
based on the MMA step, we simply define
 \beq
 \Gammabf_4 \equiv \begin{bmatrix} 0\\ 0\end{bmatrix},\quad
 F_4\equiv0,\quad
 G_4\equiv0,\quad
 R_4\equiv0.
 \eeq
It is crucial then that the finite element solver allows different parts 
of the problem to be solved in a decoupled manner, i.e., it must be 
possible to solve Eqs.~\eqref{eq:divform}-\eqref{eq:neumann} for $u_i$ 
for $i=1,2,3$ while keeping $u_4$, i.e., $\gamma$, fixed.

In \Femlab{} the nonlinear problem Eq.~\eqref{eq:Li+Mi} is solved using
damped Newton iterations~\cite{Refman}. Therefore the matrices
$\Kbf_{ij}=-\partial\Lbf_i/\partial\Ubf_j$ and
$\Nbf_{ij}=-\partial\Mbf_i/\partial\Ubf_j$ appearing in the adjoint 
problem Eq.~\eqref{eq:adjoint} are computed automatically as part of 
the solution process and can be obtained directly as \Matlab{} sparse 
matrices. They are given by
 \begin{align}
 \Kbf_{ij,nm} =& -\intO \Big(\varphi_{i,n}\Big[
  \frac{\partial F_i}{\partial u_j}\,\varphi_{j,m}
  + \frac{\partial F_i}{\partial\nablabf u_j\!\!} \cdot 
   \nablabf\varphi_{j,m}\Big]
  + \nablabf\varphi_{i,n}\cdot\Big[ 
   \frac{\partial\Gammabf_i}{\partial u_j}\,\varphi_{j,m}
  + \frac{\partial\Gammabf_i}{\partial\nablabf u_j\!\!} \cdot 
   \nablabf\varphi_{j,m}\Big]\Big)\,\id\rbf \nonumber \\
 &\quad -\intdO\varphi_{i,n} \frac{\partial G_i}{\partial u_j} \, 
  \varphi_{j,m}\,\id s \label{eq:Kmatrix}
 \end{align}
and
 \beq
 \Nbf_{ij,nm} = -\frac{\partial R_i}{\partial u_j} 
  \Big|_{\rbf_{i,n}}\,\varphi_{j,m}(\rbf_{i,n}).
 \eeq
Regarding the right-hand side vector $\partial\Phi/\partial\Ubf_i$ in
Eq.~\eqref{eq:adjoint}, notice that for a general objective as
Eq.~\eqref{eq:ABPhi}, it has the form
 \beq[gradABPhi]
 \frac{\partial\Phi}{\partial u_{i,n}\!\!\!}\,\,
  = \int_\Omega\Big(\frac{\partial A}{\partial u_i}
    + \frac{\partial A}{\partial\nablabf u_i\!\!}
     \cdot \nablabf\Big)\varphi_{i,n}\,\id\rbf
   + \int_{\partial\Omega}
    \frac{\partial B}{\partial u_i}\,\varphi_{i,n}\,\id s.
 \eeq
It is not in the spirit of a high-level finite element package to 
program the assembly of this vector by hand. In stead we employ the 
built-in assembly subroutine of \Femlab{}. We construct a copy of the 
original problem sharing the geometry, finite element mesh, and 
degree-of-freedom numbering with the original. Only we replace the 
original fields $\Gammabf_i$, $F_i$, and $G_i$ with
 \beq[adjointfields]
 \widetilde{\Gammabf}_i \equiv
  \frac{\partial A}{\partial\nablabf u_i\!\!}\, ,\quad
 \widetilde{F}_i \equiv \frac{\partial A}{\partial u_i},
 \quad\mbox{and}\quad
 \widetilde{G}_i \equiv \frac{\partial B}{\partial u_i}\,.
 \eeq
Assembling the right-hand-side vector $\widetilde{\Lbf}_i$ with
this definition yields exactly Eq.~\eqref{eq:gradABPhi}, c.f.
Eq.~\eqref{eq:Lint}. An extra convenience in \Femlab{} is that we
can rely on the built-in symbolic differentiation tools to compute
the derivatives $\partial A/\partial u_i$ etc. In order to try out
a new objective for the optimization problem, the user essentially
only needs to change the text expressions defining the quantities
$A$ and $B$.

After solving the adjoint problem Eq.~\eqref{eq:adjoint} for
$\widetilde{\Ubf}_i$ and $\widetilde{\Lambdabf}_i$ to eliminate
$\partial\Ubf_i/\partial\gammabf$ and 
$\partial\Lambdabf_i/\partial\gammabf$ for $i=1,2,3$ in 
Eq.~\eqref{eq:undesired} we can evaluate the sensitivity
 \begin{align}
 \frac{d}{d\gammabf}\Big[\Phi(\Ubf,\gammabf)\Big] &=
  \frac{\partial\Phi}{\partial\gammabf} +
  \sum_{j=1}^3\Big(\frac{\partial\Lbf_j}{\partial\gammabf}\Big)^T
   \widetilde{\Ubf}_j - \Big(\frac{\partial\Mbf_j}{\partial\gammabf}
   \Big)^T\widetilde{\Lambdabf}_j \nonumber\\
 &= \widetilde{\Lbf}_4 - \sum_{j=1}^3\Big(\Kbf_{j4}^T\widetilde{\Ubf}_j
   - \Nbf_{j4}^T\widetilde{\Lambdabf}_j\Big),\label{eq:sensitivity}
 \end{align}
where $\Kbf_{i4}=-\partial\Lbf_i/\partial\gammabf$, 
$\Nbf_{i,4} = -\partial\Mbf_i/\partial\gammabf$, and 
$\widetilde{\Lbf}_4 = \partial\Phi/\partial\gammabf$ in accordance with 
$\Ubf_4 \equiv \gammabf$.
Since the fourth variable $\gamma$ is treated on equal footing with the 
other three variables, all expressions required to compute the matrices 
$\Kbf_{i,4}$ and $\Nbf_{i,4}$ come out of the standard linearization of 
the problem. This is yet another advantage of including $\gamma$ as an 
extra dependent variable.

When dealing with a problem with a volume constraint as in
Eq.~\eqref{eq:stvol}, it is necessary to compute the derivative of
the constraint with respect to $\gammabf$,
 \beq
 \frac{\partial}{\partial\gamma_n}
  \bigg[\frac{1}{|\Omega|}\intO\gamma(\rbf)\,\id\rbf - \beta\bigg]
  = \frac{1}{|\Omega|}\intO\varphi_{n,4}(\rbf)\,\id\rbf,
 \eeq
which can be obtained by assembling $\hat{\Lbf}_4$ with
$\hat{\Gammabf}_4 \equiv \Obf$, $\hat{F}_4 \equiv 1$, and
$\hat{G}_4 \equiv 0$. In the appendix we have included a transcript of 
the code required to set up and solve the example from 
Sec.~\ref{sec:4terminal} below with \Femlab{}.
It amounts to 111 lines of code, of which the majority are spent on 
setting up the actual Navier--Stokes flow problem. Only a minor part 
goes to set up the adjoint problem and perform the sensitivity analysis.
Moreover, this part contains almost no reference to the actual physical 
problem being solved, and therefore it should apply for any multi-field 
problem expressed in the divergence form Eqs.~\eqref{eq:divform} to 
\eqref{eq:neumann} with an objective function of the form of 
Eq.~\eqref{eq:ABPhi}.
The code example employs, but does not include, a \Matlab{} 
implementation of the MMA optimization 
algorithm~\cite{Svanberg88,Svanberg02,mailto:Svanberg}.

\subsubsection{Mesh dependence and regularization techniques}
It is well known that many topology optimization problems have trouble 
with mesh dependence. E.g. in stiffness design of mechanical structures 
it often pays to replace a thick beam with two thinner beams for a given 
amount of material.
As the finite element mesh is refined, smaller and smaller features can 
be resolved and therefore appear in the optimized structure. In that 
sense the flow problem that we consider here is atypical because it is 
generally unfavorable to replace a wide channel with two narrower 
channels; hence the proof for the existence of a unique optimal solution 
with respect to minimization of the total power dissipation in 
Ref.~\cite{Borrvall}.

The problem with mesh dependence can be overcome by various 
regularization techniques based on filtering of either the design
variable $\gamma(\rbf)$ or the sensitivity
$\partial\Phi/\partial\gamma$~\cite{Bendsoe03}.
The regularization works by defining a certain length scale $r_0$ 
below which any features in $\gamma(\rbf)$ or 
$\partial\Phi/\partial\gamma$ are smeared out by the filter; 
in both cases this results in optimized structures with a minimal 
feature size $\sim r_0$ independent of the mesh refinement.
Unfortunately \Femlab{} does not come with such a filter, and hence
its implementation is an issue that has to be dealt with before our 
methodology here can be succesfully applied to problems that display 
mesh dependence.

One strategy is to implement the convolution operation of the filter 
directly~\cite{Bendsoe03}. If the computational domain is rectangular 
and discretized by square finite elements this is both efficient and 
fairly easy to program, if not one simply uses a standard filter from 
the \Matlab{} Image Processing Toolbox.
For an unstructured mesh of triangular elements the programming is more 
involved and slow in \Matlab{} due to the need to loop over the design 
variable nodes and searching the mesh for neighbouring nodes within the 
filter radius. Therefore an explicit matrix representation of the filter 
would often be preferred~\cite{tofilter}.

Another possible strategy is to solve an artificial diffusion problem 
for the design variable $\gamma(\rbf)$ over some period in "time"
$\Delta t=r_0^2/k$ where $k$ is the "diffusion" constant.
The diffusion equation could be included into the fields of
Eq.~\eqref{eq:adjointfields} that are otherwise unused, and the "time" 
evolution solved using the built-in timestepper in \Femlab{}.
This procedure is equivalent to the action of a filter with Gaussian 
kernel of width $r_0$, and it conserves the total amount of material 
during the filter action.
The same approach could be used to smooth out the sensitivity. However, 
because $\partial\Phi/\partial\gamma_n$ is sensitive to the local 
element size one would need to rescale it with 
$\int_\Omega\phi_{4,n}(\rbf)\,\id\rbf$ before application of the
filter -- actually this is true for any filter acting on
$\partial\Phi/\partial\gamma$ whenever the mesh is irregular and
$\int_\Omega\phi_{4,n}(\rbf)\,\id\rbf$ not constant for all $n$.

The major disadvantage of this strategy is that it involves solving a 
time evolution problem in each design iteration which could easily turn 
out to be the most time-consuming step.
Alternatively the timestepping algorithm could be implemented by
hand, e.g., deciding on the Crank-Nicholson algorithm with a fixed
stepsize $\delta t\leq\Delta t$. The mass and stiffness matrices
for the diffusion problem can be obtained from \Femlab{}, and the
corresponding iteration matrix need only be factorized once for the 
given stepsize and could thus be reused in all subsequent design 
iterations, making this approach relatively cheap, although more 
cumbersome than using the built-in timestepper.

\subsubsection{Large-scale problems}
For large scale problems and three dimensional modeling it is often
necessary to resort to iterative linear solvers because the memory
requirements of a direct matrix factorization becomes prohibitive.
In that case the strategy we have outlined here of obtaining the 
$\Kbf$ and $\Nbf$ matrices directly as sparse matrices in \Matlab{} 
and simply transposing $\Kbf$ before the solution of the adjoint 
problem may not be practical.
Alternatively, if the original physical problem is expressed in
divergence form then the \Femlab{} representation of that problem
contains the symbolic derivatives of $\Gammabf_i$, $F_i$, and $G_i$
appearing in Eq.~\eqref{eq:Kmatrix}.
These fields can be transposed and set in the auxiliary copy of
the original problem such that it effectively defines
$\tilde{\Kbf}_{ij} = \Kbf_{ji}^T$, while retaining the
definitions in Eq.~\eqref{eq:adjointfields} for the right-hand-side
vector $\tilde{\Lbf}_i$.
Then the adjoint problem Eq.~\eqref{eq:adjoint} can be solved
\emph{without} direct handling of the matrices in \Matlab{}, and
using the same iterative solver algorithm as would be employed for
the original physical problem.
Ultimately we still require an explicit representation of the 
matrices $\Kbf_{i,4}$ and $\Nbf_{i,4}$ to evaluate the sensitivity
$\partial\Phi/\partial\gammabf$ in Eq.~\eqref{eq:sensitivity}.

From our point of view the major advantage of using \Femlab{} in
its present stage of development for topology optimization is not
in solving large scale problems, though, but rather in the ease of
implementation and the ability to handle problems with coupling
between several physical processes.

\section{NUMERICAL EXAMPLES}
\label{sec:examples} In this section we present our results for
topology optimization of Navier--Stokes flow for two particular
model systems that we have studied. These systems have been chosen
because they illustrate the dependence of the solution on the two
dimensionless numbers $Re$ and $Da$, measuring the importance of
the inertia of the fluid and the permeability of the porous
medium, respectively, relative to viscosity. Moreover we discuss
the dependence of the solution on the initial condition for the
material distribution.

For simplicity and clarity we have chosen to consider only 
two-dimensional model systems.
We note that the dimensionality of the problems has no fundamental 
consequence for the method and the numerics, but only affects 
computer memory requirements and the demand for CPU time.
Our 2D examples can therefore be viewed as idealized test cases
for our implementation of topology optimization.
Yet, the 2D models are not entirely of academic interest only as 
they represent two limits of actual 3D systems.
Due to planar process technology many contemporary lab-on-a-chip 
systems have a flat geometry with typical channel heights of about 
10~$\mu$m and widths of 1~mm, i.e., an aspect ratio of 
1:100~\cite{Geschke03}.
One limit is the case where the channel width is constant and the
channel substrate and lid are patterned with a profile that is 
translation invariant in the transverse (width) direction.
In the limit of infinitely wide channels the 2D-flow in the plane 
perpendicular to the width-direction is an exact solution, while
it remains an excellent approximation in a 1:100 aspect ratio
channel. This is the model system we have adopted for the 
numerical examples in the present work.
The other important limit is when the channel width is not constant, 
but the channel height is sufficiently slowly varying
that the vertical component of the fluid velocity can be neglected.
Then writing the Navier--Stokes equation for the velocity 
averaged in the vertical (height) direction, the out-of-plane
shear imposed by the channel substrate and lid gives rise to an
absorption term~$-\alpha\vbf$. This approach was studied by
Borrvall and Petersson~\cite{Borrvall}, see also the footnote in 
Sec.~\ref{sec:GovEqs}.
Thus, if one is interested in optimizing the height-averaged flow
field in a flat channel the 2D model is sufficient.

When solving the Navier--Stokes flow problem we use the standard
direct linear solver in \Femlab{} in the Newton iterations.
Typically we have around 6000 elements in the mesh, corresponding
to 30000 degrees-of-freedom. The constrained optimization problem
is solved using a \Matlab{} implementation of the MMA algorithm
kindly provided by K. Svanberg~\cite{Svanberg88,mailto:Svanberg},
except that we modified the code to use the globally convergent
scheme described in Ref.~\cite{Svanberg02}.
The example script included in the appendix employs only the basic
algorithm \texttt{mmasub}, though.
The design iterations are stopped when the maximal change in the design 
field is $\|\gamma^{(k+1)}-\gamma^{(k)}\|_\infty\leq 0.01$, at which 
point we typically have $|\Phi^{(k+1)}-\Phi^{(k)}|< 10^{-5}$.

\subsection{Example: a channel with reverse flow}
\label{sec:reverseflow} Our first numerical example deals with the design 
of a structure that at a particular point inside a long straight channel 
can guide the flow in the opposite direction of the applied pressure drop.
The corresponding problem with a prescribed flow rate was first suggested
and investigated by A.~Gersborg-Hansen~\cite{GersborgHansen03}.
We elaborate on it here to illustrate the importance of the choice of
permeability for the porous medium.

\begin{figure}
\centerline{\includegraphics[width=0.6\textwidth]{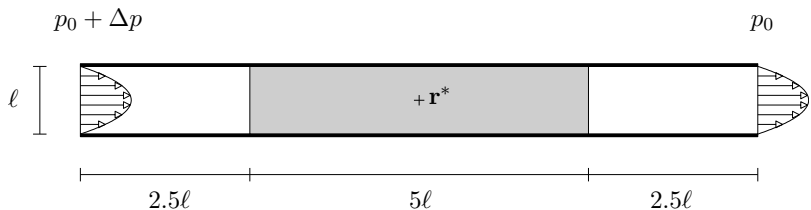}}
\caption{\label{fig:Sgeom}
Computational domain for the reverse flow example. The design domain 
(gray) has length $5\ell$ and height $\ell$, and the fluid enters and 
leaves the design domain through leads of length $2.5\ell$.
The boundary conditions prescribe a pressure drop of $\Delta p$ across 
the system, and the design objective is to reverse the flow direction
at the point $\rbf^*$ at center of the channel.}
\end{figure}

The computational domain is shown in Fig.~\ref{fig:Sgeom}. It consists 
of a long straight channel of height $\ell$ and length $L=10\ell$; the 
actual design domain, inside which the porous material is distributed, 
is limited to the central part of length $5\ell$. The boundary conditions 
prescribe a pressure drop of $\Delta p$ from the inlet (left) to the 
outlet (right), and no-slip for the fluid on the channel side walls.

The optimization problem is stated as a minimization of the horizontal 
fluid velocity at the point $\rbf^*$ at the center of the channel, 
i.e., the design objective is
 \beq
 \Phi = v_1(\rbf^*).
 \eeq
In terms of the general objective Eq.~\eqref{eq:ABPhi} this is obtained 
with $A \equiv v_1(\rbf)\delta(\rbf-\rbf^*)$ and $B \equiv 0$.
There is no explicit need for a volume constraint because neither of the
extreme solutions of completely filled or empty can be optimal.
When the design domain is completely filled with porous material
we expect a flat flow profile with magnitude below 
$\Delta p\,/(5\ell\alpha_{\max})$. In the other extreme case when the
channel is completely devoid of porous material the solution is simply 
a parabolic Poiseuille profile with maximum
 \beq[vmax]
 v_0 = \frac{\eta}{8\ell^2}\frac{\Delta p}{L}.
 \eeq
However, a structure that \emph{reverses} the flow such that $v_1(\rbf^*)$
becomes negative will be superior to both these extreme cases in the sense 
of minimizing $\Phi$.

\subsubsection{Reverse flow in the Stokes limit, $Re = 0$}
\label{sec:ReverseFlowStokes} We first consider the Stokes flow limit 
of small $\Delta p$ where the inertial term becomes neglible. The problem 
is then linear and the solution is characterized by a single dimensionless 
parameter, namely the Darcy number $Da$, Eq.~\eqref{eq:darcynumber}. 
We have solved the topology optimization problem for different values of 
$Da$. The initial condition for the material distribution was
$\gamma^{(0)}=1$, and the parameter $q$ determining the shape of
$\alpha(\gamma)$ in Eq.~\eqref{eq:alpha} was set to $q=0.1$.
Anticipating that the structural details close to $\rbf^*$ should
be more important than those further away we chose a non-uniform
finite element mesh with increased resolution around $\rbf^*$.

\begin{figure}
\centerline{\includegraphics[width=0.75\textwidth]{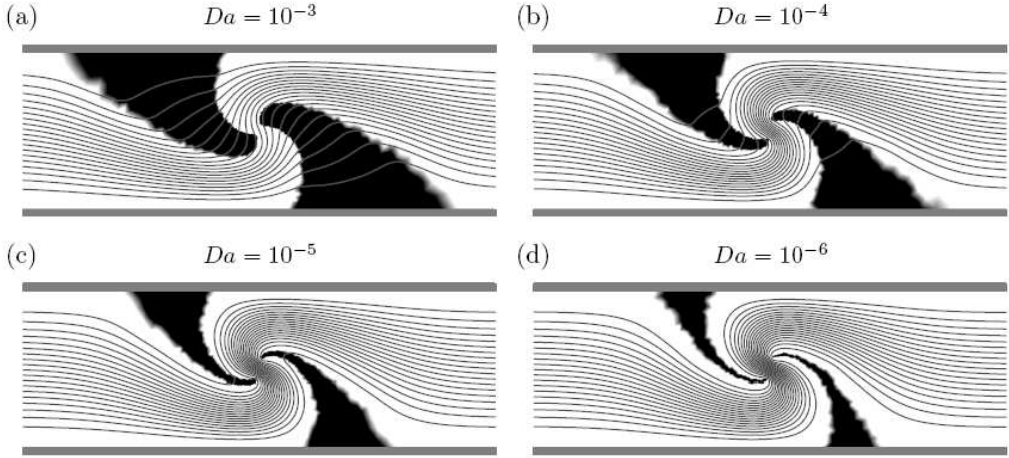}}
\caption{\label{fig:Sshape0}
Optimized structures (black) and streamlines at 5\% intervals for 
Stokes flow ($Re=0$) at Darcy numbers decreasing from $10^{-3}$ to 
$10^{-6}$.
Only the central part of length $3\ell$ of the design domain is shown.
The structures consist of two barriers defining an $S$-shaped channel
that reverses the flow at the central point $\rbf^*$.
As the Darcy number is decreased, the optimized structures become
thinner and less permeable.}
\end{figure}

Fig.~\ref{fig:Sshape0} shows the optimal structures obtained for
$Da=10^{-3}$, $10^{-4}$, $10^{-5}$, and $10^{-6}$.
They all consist of two barriers defining an $S$-shaped channel that 
guides the fluid in the reverse direction of the applied pressure drop.
At $Da=10^{-3}$ the two barriers are rather thick but leaky with with 
almost all the streamlines penetrating them; as the Darcy number is 
decreased the optimal structures become thinner and less penetrable.
This result can be interpreted as a trade-off between having either 
thick barriers or wide channels. Thick barriers are necessary to force 
the fluid into the $S$-turn, while at the same time the open channel 
should be as wide as possible in order to minimize the hydraulic 
resistance and maximize the fluid flow at the prescribed pressure drop.

Notice that if we had chosen to prescribe the flow rate through the 
device rather than the pressure drop, then the optimal solution would 
have been somewhat different. When the flow rate is prescribed, it pays 
to make the gap between the barriers very small and the barriers very 
thick in order to force the fixed amount of fluid flow through the 
narrow contraction.
The optimal structure is therefore one with a very large hydraulic 
resistance. In Ref.~\cite{GersborgHansen03} this problem was circumvented 
by adding a constraint on the maximal power dissipation allowed at the 
given flow rate.

\begin{figure}
\centerline{\includegraphics[width=0.45\textwidth]{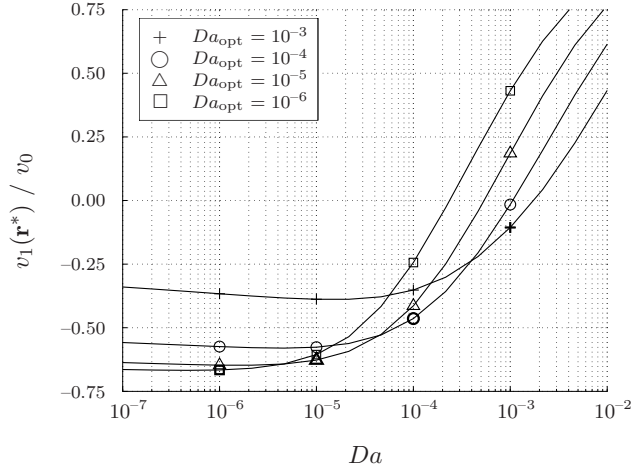}}
\caption{\label{fig:chkamx}
Comparing the performance of the structures from Fig.~\ref{fig:Sshape0}
optimized at $Da_\textrm{opt}$ for different values of $Da$.
The objective $v_1(\rbf^*)$ is normalized with the velocity in an 
empty channel, $v_0$, c.f.~Eq.~\eqref{eq:vmax}.}
\end{figure}

In order to validate the optimality of the structures computed by
the topology optimization we do as follows:
For each of the optimized structures from Fig.~\ref{fig:Sshape0}
we freeze the material distribution and solve the flow problem for
a range of Darcy numbers.
The resulting family of curves for $v_1(\rbf^*)$ vs. $Da$ is shown 
in Fig.~\ref{fig:chkamx} where it is seen that each of the four 
structures from Fig.~\ref{fig:Sshape0} do indeed perform better
in minimizing $v_1(\rbf^*)$ than the others at the value of $Da$
for which they are optimized.

For $Da\lesssim 10^{-5}$ the optimal value of $v_1(\rbf^*)$ tends
to saturate because the thin barriers are then almost completely 
impermeable and the open channel cannot get much wider.
In this limit the thickness of the optimized barrier structures
approach the mesh resolution as seen in Fig.~\ref{fig:Sshape0}(d).
When the optimal barrier thickness gets below the mesh size we have 
observed the appearance of artificial local optima for the barrier 
structure. The problem is that the thin barriers cannot continuously 
deform into another position without going through an intermediate 
structure with barriers that are thicker by at least one mesh element. 
Depending on the initial condition, the optimization algorithm can
therefore end up with a sub-optimal structure.
We have tried to work around this problem by decreasing the value of 
$q$ in order to smear out the solid/void interfaces and thus reduce 
the cost of going through the intermediate structure.
This did not work out well; the reason may be that the smearing property 
of a convex $\alpha(\gamma)$ was derived for the objective of minimizing 
the power dissipation subject to a volume constraint. In the present 
example we are dealing with a different objective and have no volume 
constraint.
However, when the barrier structures are resolved with at least a few
elements across them the artificial local optima tend to be 
insignificant. Thus the problem can be avoided by choosing a
sufficiently fine mesh, or by adaptively refining the mesh at the 
solid/void interfaces.

Returning to Fig.~\ref{fig:chkamx} we notice that as $Da$ increases
all the structures perform poorly in minimizing $v_1(\rbf^*)$, as they 
all approach $v_0$.
Extrapolating this trend one might suspect that the $S$-turn topology 
will cease to be optimal somewhere above $Da=10^{-3}$ simply because the
porous material becomes too permable to make reversal of the flow
direction possible.
\begin{figure}
\centerline{\includegraphics[width=0.375\textwidth]{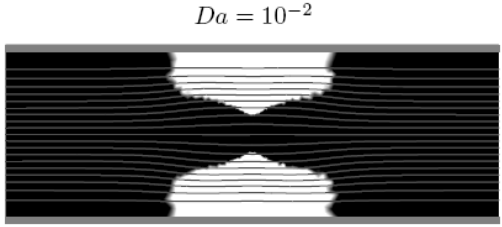}}
\caption{\label{fig:Hshape0}
Optimized structure (black) and streamlines for Stokes flow at 
$Da=10^{-2}$; only the central part of length $3\ell$.
The design domain is completely filled with porous material, except
immediately above and below $\rbf^*$ where two empty regions emerge.
These voids divert the flow away from $\rbf^*$, resulting in a low 
velocity $v_1(\rbf^*)=0.1 v_0$.}
\end{figure}
We have tested this hypothesis by performing an optimization at
$Da=10^{-2}$, resulting in the structure shown in 
Fig.~\ref{fig:Hshape0} where the value of the objective is
$v_1(\rbf^*)=0.1 v_0$. It is seen to display a different topology
from those of Fig.~\ref{fig:Sshape0}, with the design domain is
almost completely filled with porous material blocking the flow
through the channel. Only immediately above and below the point
$\rbf^*$ we see two empty regions emerging that act guide the
flow away from $\rbf^*$.

Actually, in all four cases from Fig.~\ref{fig:Sshape0}, starting
from an empty channel the design iterations initially converge
towards a symmetric structure blocking the flow like that in
Fig.~\ref{fig:Hshape0}. However at a certain point in the
iterations an asymmetry in the horizontal plane is excited and the
structure quickly changes to the two-barrier $S$-geometry. Whether
the optimization converge to an $S$- or an inverted $S$-turn
depends how the asymmetry is excited from numerical noise or
irregularity in the finite element mesh; in fact the structure in
Fig.~\ref{fig:Sshape0}(b) originally came out as an inverted $S$
but was mirrored by hand before plotting it to facilitate
comparision with the three other structures.

\subsubsection{Reverse flow at finite Reynolds number}
We now consider flow at finite Reynolds number, characterized by
the two dimensionless numbers $Re$ and $Da$. The geometry and
boundary conditions remain unchanged, for convenience we introduce
a non-dimensional pressure drop $\Delta\tilde{p} = \Delta
p\,\rho\,\ell^2/\eta^2$, and finally we fix the Darcy number at
$Da=10^{-5}$. We note from Fig.~\ref{fig:Sshape0} 
that this Darcy number allows some but not much fluid to
penetrate the walls. We have nevertheless chosen this Darcy number
for practical reasons, as the walls are "solid" enough and a lower
value (more "solid" wall) would increase the calculation time.

\begin{figure}
\includegraphics[width=0.75\textwidth]{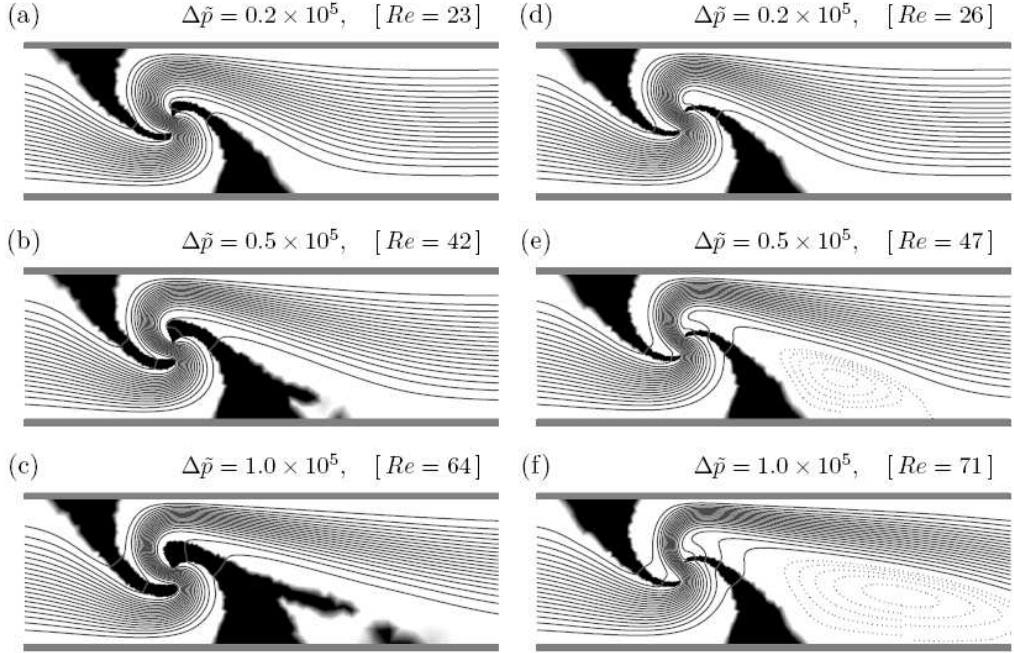}
\caption{\label{fig:Sshapef}
Optimized structures (black) and streamlines for Navier--Stokes flow;
only a part of length $3.25\ell$ near the center of the channel is shown.
Panel (a)-(c) to the left show the optimized structures for different 
values of the control parameter 
$\Delta\tilde{p} = \Delta p\,\rho\ell^2/\eta^2$.
For comparision the flow field when the optimized structure from
Fig.~\ref{fig:Sshape0}(c) is frozen and exposed to the elevated 
pressure drops is shown in panel (d)-(f) to the right.
The Reynolds number is defined as $Re = \rho\,\ell\, v_{\max}/\eta$ 
where $v_{\max}$ is the maximal velocity measured at the inlet;
note that for a particular value of $\Delta\tilde{p}$, the Reynolds 
number is not fixed but differs slightly between left and right column.}
\end{figure}

We have solved the topology optimization problem for different values of
$\Delta\tilde{p}$, always using an empty channel as initial condition.
The results are shown in Fig.~\ref{fig:Sshapef}(a)-(c) for
$\Delta\tilde{p}=0.2,0.5,$ and $1.0\times10^5$, where only a few 
streamlines are seen to penetrate the barriers.
For comparision we also consider the flow field obtained when the 
structure optimized for Stokes flow at $Da=10^{-5}$ is frozen and 
exposed to the three different elevated pressure drops. This is shown in
Fig.~\mbox{\ref{fig:Sshapef}(d)-(f)}: As the pressure drop is increased, 
more and more streamlines penetrate the barriers.
Moreover we find a recirculation region emerging behind the second barrier
which reduces the pressure drop over the neck between the barriers.

Returning to Fig.~\ref{fig:Sshapef}(a)-(c), we find that the structures
that have been optimized for the corresponding pressure drops are 
generally thicker than that optimized for Stokes flow, which reduces 
number of streamlines penetrating them.
Also a beak-like tip grows on the second barrier that acts to bend the 
fluid stream down. Finally, on the back of the second barrier a wing- or 
spoiler-like structure appears that removes the recirculation.

In summary, our first example has demonstrated that our implementation 
of topology optimization works, but that the optimal design and 
performance may depend strongly on the choice of the Darcy number. 
In particular, the zero $Da$ limit solution contains zero thickness and 
yet impermeable barriers deflecting the fluid. In order to approximate 
this solution at finite $Da$ and on a finite resolution mesh it is 
important to choose the Darcy number small enough that even thin barriers 
can be almost impermeable, but large enough to avoid difficulties with
artificial local optima in the discretized problem when the barrier 
thickness decreases below the mesh resolution.

\subsection{Example: a four-terminal device}
\label{sec:4terminal} Our second numerical example deals with 
minimization of the power dissipation in a four-terminal device subject 
to a volume constraint. The problem is found to exhibit a discrete change 
in optimal topology driven by the inertial term. The four-terminal
device is related to one considered by Borrvall and Petersson for
Stokes flow in Ref.~\cite{Borrvall}; the present example demonstrates
that the optimization algorithm has difficulties in finding
the optimal topology when there are two strong candidates for the
global optimum.

\begin{figure}
\centerline{\includegraphics[width=0.4125\textwidth]{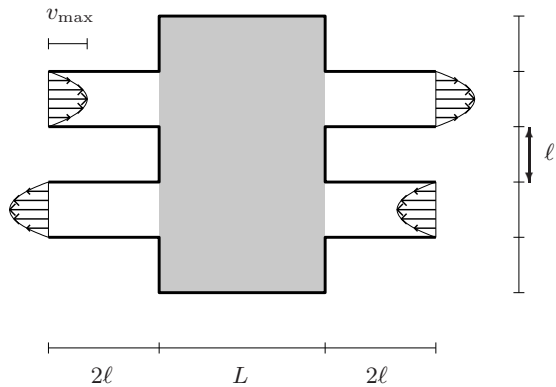}}
\caption{\label{fig:setup}
Schematic illustration of the four-terminal device. Two inlet and two 
outlet leads (white areas) of height $\ell$ and length $2\ell$ are 
attached to the design domain (gray) of height $5\ell$ and length $L$.
The flow is characterized by the Reynolds number 
$Re=\rho\ell v_{\max}/\eta$, where $v_{\max}$ is the maximal velocity 
at the inlets.}
\end{figure}

The computational domain, shown in Fig.~\ref{fig:setup}, consists of a 
rectangular design domain (gray) to which two inlet and two outlet leads 
(white) are attached symmetrically.
The boundary conditions prescribe parabolic profiles for the flow at
the inlets, zero pressure and normal flow at the outlets, and no-slip 
on all other external boundaries.
Choosing the height $\ell$ of the leads as our characteristic length 
scale, we define the Reynolds number as $Re=\rho\ell v_{\max}/\eta$, 
where $v_{\max}$ is the maximal velocity at the inlets.
The Darcy number is fixed at $Da=10^{-4}$ to obtain reasonably small 
leakage through the porous walls.

The optimization problem is stated as a minimization of the total power
dissipation inside the computational domain, given by 
Eq.~\eqref{eq:power}, subject to the constraint that at most a fraction
$\beta=0.4$ of the design domain should be without porous material,
c.f. Eq.~\eqref{eq:stvol}.

\begin{figure}
\centerline{\includegraphics[width=0.75\textwidth]{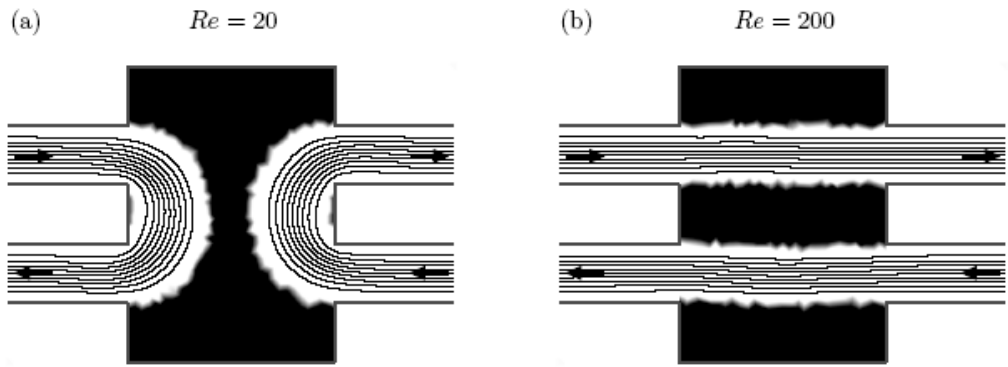}}
\caption{\label{fig:UIturns}
Optimal structures (black) and streamlines at 10\% intervals for the 
four-terminal device at Reynolds number $Re=20$ and 200, respectively,
in a geometry with $L=3.5\ell$.}
\end{figure}

Fig.~\ref{fig:UIturns}(a) and (b) shows the two optimal structures
obtained for $Re=20$ and 200, respectively, in a geometry with
$L=3.5\ell$. At $Re=20$ the optimal structure turns out to be a
pair of $U$-turns connecting the inlets to the outlets on the same
side of the design domain, while at $Re=200$ the optimal structure
is a pair of parallel channels. In order to minimize the power
dissipation at low $Re$, the channel segments should be as short
and as wide as possible, which favors the $U$-turns in
Fig.~\ref{fig:UIturns}(a). However, as the Reynolds number is
increased, the cost of bending the fluid stream grows. When
inertia dominates, larger velocity gradients appear in the long
"outer lane" of the $U$-turn. This increases the dissipation
compared to low $Re$, where more fluid flows in the shorter "inner
lane". At a certain point it will exceed the dissipation in the
parallel channels solution
 \beq[Phi0]
 \Phi_0 = \frac{96}{9}\Big(4+\frac{L}{\ell}\Big)\,\eta v_{\max}^2,
 \eeq
as estimated from Poiseuille flow in two straight channels, each of 
length $L+4\ell$ and height $\ell$. This number is independent of 
inertia due to translation symmetry, and we use $\Phi_0$ as a natural 
unit of power dissipation (per unit length in the third dimension) in 
the following.

Clearly the Reynolds number at which the transition between the two 
classes of solutions occurs will depend strongly on the ratio $L/\ell$.
For short lengths $L\lesssim 2\ell$ the parallel channels solution is 
expected to be optimal at all $Re$, whereas for long lengths 
$L\gtrsim 3\ell$ the $U$-turn solution should be significantly better 
than the parallel channels solution at low $Re$.

\subsubsection{Dependence on the Reynolds number}
In the following we investigate more closely the transition between 
the $U$-turns and the parallel channels solution as a function of the
Reynolds number for the particular geometry $L=3\ell$.
The topology optimization problem is solved for different $Re$ in the 
range 0 to 200, using a homogeneous material distribution
$\gamma^{(0)} = 0.4$ as initial condition.
For the parameter $q$ determining the shape of $\alpha(\gamma)$ in
Eq.~\eqref{eq:alpha} we use a two-step solution procedure as suggested 
in Ref.~\cite{Borrvall}.
First the problem is solved with $q=0.01$ in order to obtain a solution
with slightly smeared-out solid/void interfaces.
Next this material distribution is used as initial guess for an 
optimization with $q=0.1$ which generates fully discrete solid/void 
interfaces at the resolution of our finite element mesh.

\begin{figure}
\centerline{\includegraphics[width=0.7275\textwidth]{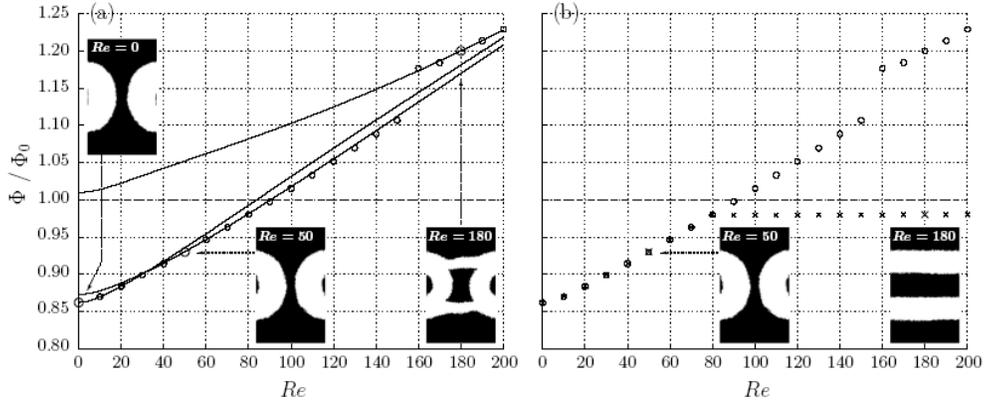}}
\caption{\label{fig:chkL30f104}
Power dissipation $\Phi$ in structures optimized for different
Reynolds numbers; normalized with the Poiseuille flow result $\Phi_0$
(dashed line).
(a)~Markers $(\circ)$ show results when $\gamma^{(0)}=0.4$ is used
as initial condition, failing to find the optimal solution for $Re>90$.
Full lines show the performance of the structures optimized
at $Re=0$, 50, and 180, when evaluated at different Reynolds numbers.
As expected, all points fall on or below the full lines, except the
hybrid solutions for $Re\geq160$.
(b)~Comparision between the two different initial conditions
$\gamma^{(0)}=0.4$ $(\circ)$ and $\gamma^{(0)}=1$ $(\times)$, showing 
the success of the empty channel initial condition in finding the 
optimal solution.
The crosses $(\times)$ fall slightly below $\Phi/\Phi_0=1$ due to 
leakage through the porous walls (see the text).}
\end{figure}

Fig.~\ref{fig:chkL30f104}(a) shows the result for the normalized power
dissipation $\Phi/\Phi_0$ obtained as a function of $Re$.
At low Reynolds numbers the optimized solutions correctly come out as
$U$-turns with a power dissipation $\Phi$ that is clearly less than
$\Phi_0$.
However, at high Reynolds numbers $Re > 90$ the method fails because
the optimized solutions continue to come out as $U$-turns even though
this yields $\Phi/\Phi_0>1$.
For $Re\geq 160$ the solution jumps from the simple $U$-turns to a
hybrid structure, as shown in the inset.
The full lines in Fig.~\ref{fig:chkL30f104}(a) show the result when the
material distributions optimized for $Re=0$, 50, and 180, respectively, 
are frozen and the power dissipation evaluated at different $Re$.
It is seen that the optimized solutions, marked $(\circ)$, all fall on 
or below the full lines which confirms that they are indeed superior to 
the other solutions of the $U$-turn family.
This also holds for $Re>90$, except for the hybrid structures at 
$Re\geq 160$, that are actually inferior to the $U$-turns.
Moreover, at $Re=160$ the optimized solution falls slightly above that 
optimized at $Re=180$.
This could be an indication that the hybrid structures are not local 
optima in design space after all, but rather a very narrow saddle point
that the optimization algorithm has a hard time getting away from.

The difficulty is that the two families of solutions, the $U$-turns and
the parallel channels, are both deep local minima for the power 
dissipation in design space.
Using $\gamma^{(0)} = 0.4$ as initial condition, the initial 
permeability is everywhere very low, such that the porous friction 
almost completely dominates the inertia and viscous friction in the 
fluid. Therefore the iteration path in design space is biased towards 
low Reynolds numbers and the $U$-turns solution.

In order to circumvent this problem we have tried using a completely 
empty design domain with $\gamma^{(0)}=1$ as initial condition.
This should remove the bias towards the $U$-turns and allow the 
optimization algorithm to take inertia into account from iteration one.
The result is shown in Fig.~\ref{fig:chkL30f104}(b).
For $Re\leq80$ the solutions are still $U$-turns, whereas for $Re\geq 90$
they come out as parallel channels.
Notice that $\Phi/\Phi_0$ for the parallel channels solution is actually
slightly less than unity, namely 0.98.
This is due to a small amount of fluid seeping through the porous walls
defining the device, which lowers the hydraulic resistance compared to 
the Poiseuille flow result derived for solid walls.%
\footnote{The flow in a straight channel of height $\ell$ bounded by
two porous walls of thickness $\ell$ can easily be found analytically.
At $Da=10^{-4}$ the hydraulic resistance of this system is 94\% of that
for a channel of height $\ell$ bounded by solid walls, and it approaches 
this zero $Da$ limit only as $\sqrt{Da}$.
When $L=3\ell$ we therefore expect a power dissipation
$\Phi/\Phi_0 = (3\times 0.94 + 4)/7 = 0.97$ for the parallel channels
solution, including the leads. This is close to the observed 0.98.}

Strictly speaking the initial condition $\gamma^{(0)}=1$ is not a 
feasible solution because it violates the volume constraint that at 
least a fraction $1-\beta=0.6$ of the design domain should be filled 
with porous material.
However, the MMA optimization algorithm penalizes this and reaches a 
feasible solution after a few iterations.
This is controlled by choosing a penalty parameter.
If the penalty for violating the constraint is small, the material
is added slowly and only where it does not disturb the flow much.
If the penalty is large, the material is added quickly and almost
homogeneously until the constraint is satisfied.
The succesful result from Fig.~\ref{fig:chkL30f104}(b) was obtained with a
moderate penalty.
\begin{figure}
\centerline{\includegraphics[width=0.375\textwidth]{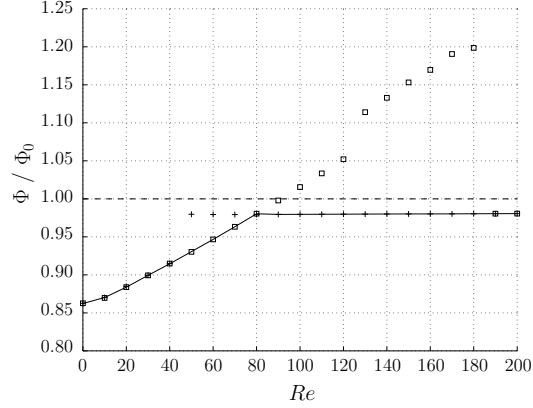}}
\caption{\label{fig:chkL30f10}
Comparision between structures optimized with the initially
non-feasible material distribution $\gamma^{(0)}=1$ for different
penalty parameters in the MMA optimization algorithm, revealing the
difficulty in choosing the condition for finding the global optimum.
Full line: the successful result from Fig.~\ref{fig:chkL30f104}(b) 
with moderate penalty;
$(+)$:~lower penalty yielding wrong result for $40<Re<80$;
$(\square)$~higher penalty yielding wrong results for $80<Re<190$.}
\end{figure}
In Fig.~\ref{fig:chkL30f10} this is compared with results for smaller
and larger penalty parameters, respectively.
The figure shows that with the small penalization, the solution jumps
to the parallel channels already at $Re=50$ which is not optimal.
For the large penalization, the solution does not jump until
$Re\geq190$. Also we observe hybrid structures similiar to those in
Fig.~\ref{fig:chkL30f104}(a) for $Re\geq130$.
We have thus not full control over the convergence towards the global
optimum.

\subsubsection{Discussion of problems with local optima}
\begin{figure}
\centerline{\includegraphics[width=0.6\textwidth]{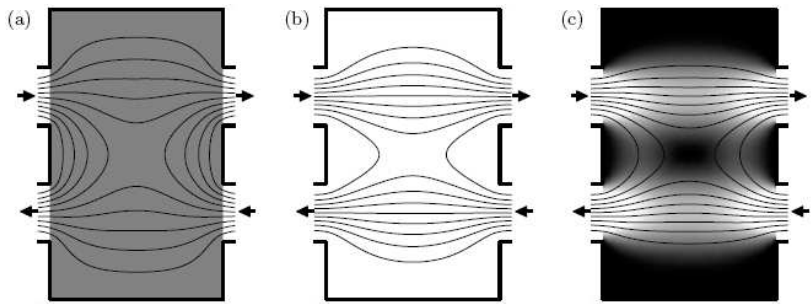}}
\caption{\label{fig:iteration1}
Flow distributions at $Re=0$, $L=3\ell$ and $q=0.01$.
(a)~Initial design field $\gamma^{(0)}=0.4$.
(b)~Initial design field $\gamma^{(0)}=0.4$.
(c)~The optimal design field $\gamma^{(*)}$ obtained at
$Da=10^{-2}$.}
\end{figure}

Further insight into the problem of local versus global optima is 
gained by inspecting the flow field in the initial material 
distribution $\gamma^{(0)}$.
This is shown in Fig.~\ref{fig:iteration1} for the Stokes flow limit, 
$Re=0$.
The streamlines are drawn as 10\% contours of the streamfunction, 
and Fig.~\ref{fig:iteration1}(a) shows that for $\gamma^{(0)}=0.4$ 
the streamline density is largest between the two leads on the same 
side of the design domain.
Based on the sensitivity $\partial\Phi/\partial\gamma$ the 
optimization algoritm therefore decides to remove material from these 
strong-flow regions in order to reduce the porous friction.
The iteration path in design space is therefore biased towards the 
$U$-turn solution. 
This remains true even at finite Reynolds numbers as long as the 
porous friction initially dominates inertia.

Fig.~\ref{fig:iteration1}(b) shows that when $\gamma^{(0)}=1$ the 
streamline density is largest between the leads on the opposite side 
of the design domain.
Because the volume constraint is violated the optimization algorithm has 
to place material somewhere, which it does in the weak-flow regions.
The solution is therefore biased towards the parallel channels.
Indeed if the penalty is chosen very small, the optimized solution comes 
out as parallel channels even for Stokes flow at $L=3\ell$, which is far 
from optimal.
When the penalty is larger and the material is added faster, we move 
away from this adiabatic solution and closer to the situation for 
$\gamma^{(0)}=0.4$.

The additional complexity associated with making a proper choice
of the penalty parameter is somewhat inconvenient. We have therefore 
attempted to construct a more convex problem by increasing the initial 
permeability. This can be done either by increasing the Darcy number, 
or by decreasing the parameter $q$, c.f. Eq.~\eqref{eq:alpha}.
Fig.~\ref{fig:iteration1}(c) shows the optimal solution $\gamma^{(*)}$ 
obtained for $Da=10^{-2}$ and $q=0.01$ at $Re=0$. At this level the 
problem is convex because the solution is independent of the initial 
condition. Using this material distribution as initial guess and 
gradually decreasing the permeability to $Da=10^{-4}$ and $q=0.1$ we 
correctly end up in the $U$-turn solution. However, it is evident from
Fig.~\ref{fig:iteration1}(c) that $\gamma^{(*)}$ has a fair amount
of parallel channels nature. Using the same procedure of gradually 
decreasing the permeability at higher Reynolds numbers therefore result 
in a transition to the parallel channels solution already for $Re\geq30$, 
which is not optimal. Moreover, when the Reynolds number is increased 
and the inertia starts to play in, the system tends to loose convexity 
even at the initial high permeability.

In summary, the topology optimization has difficulties in finding the 
global optimum for the problem. There are two strong candidates for the 
optimal structure, and the solution found is sensitive to the initial 
condition for the material distribution.
Using an empty channel as initial condition, the method is able to find 
the correct solution for all Reynolds numbers. However, this successful 
result depends on a particular choice of the penalty parameter in the 
MMA algorithm. By using a high initial permeability of the porous medium,
it is possible to convexify the problem at low Reynolds numbers, but 
continuation of this solution to the desired low permeability does not 
generally lead to the global minimum of the non-convex problem.

In the original paper Ref.~\cite{Borrvall} it was argued that in
Stokes flow the true optimal design should be rather insensitive
to the choice of the Darcy number, although the dissipated power
may deviate quite a lot from the zero $Da$ limit. In our work we
have observed that the actual solution found by the topology
optimization may depend a great deal on the choice of the Darcy
number, whereas the dissipated power should approach the zero $Da$
limit roughly as~$\sqrt{Da}$.

\section{CONCLUSION}
Based on the work of Borrvall and Petersson~\cite{Borrvall} we
have extended the topology optimization of fluid networks to cover
the full incompressible Navier--Stokes equations in steady-state.
Our implementation of the method is based of the commercial finite
element package \Femlab{}, which reduces the programming effort
required to a minimum. Formulating the problem in terms of a general 
integral-type objective function and expressing the governing 
equations in divergence form makes the implementation very compact 
and transparent. Moreover the code for performing the sensitivity 
analysis should remain almost the same for any problem expressed in 
this way, whereas that required for describing the physical problem 
of course changes. Topology optimization of multi-field problems can 
therefore be dealt with almost as easy as a single realization of 
the underlying physical problem.

We would like to mention that our methodology is not as such restricted 
to the (large) class of physical problems that can be expressed in 
divergence form.
\Femlab{} also allows problems to be stated directly in weak form, 
e.g., for systems with dynamics at the boundaries. This does in fact 
not invalidate the sensitivity analysis worked out in
Sec.~\ref{sec:discretization}, since this analysis only relies on
the basic structure of the discretized nonlinear problem and the 
availability of the Jacobian matrix. It is therefore possible to 
apply our methodology to even larger classes of physical problems 
than the ones comprised by the divergence form.

Our implementation of topology optimization has been tested on two
fluidics examples in~2D, both illustrating the influence of different 
quantities and conditions on the efficiency of the optimization method.

The first example, a channel with reversed flow, illustrates the 
influence of the Reynolds number $Re$ and the Darcy number $Da$ on
the solutions. We have shown that the choice of $Da$ has a strong
impact on the solution when the structure contains barriers to deflect 
the fluid stream.

The second example, minimization of the power dissipation in a
four-terminal device, reveals the problems of determining the global
minimum when two strong minima are competing. This problem is
highly non-convex, and we have shown that the solution depends on
the initial condition. For an initial homogeneous material distribution, 
the porous friction dominates and the solution does not come out as
the global optimum in all cases. Using an empty channel as the initial 
state, inertia plays a role from the beginning, and better results
can be obtained.
However, this initial condition in fact violates the volume constraint,
and the part of the optimization routine correcting this depends on a
penalty factor. Unfortunately, the particular value chosen for this
factor strongly influences the results.
Increasing the Darcy number makes the problem more convex, but
continuation from large to small $Da$, i.e., from high to low
permeability of the porous material, does not generally end up in the
global optimum.

In conclusion, we have shown that our implementation of topology
optimization is a useful tool for designing fluidic devices.

\section{ACKNOWLEDGEMENTS}
We are grateful to Ole Sigmund and Allan Gersborg-Hansen for 
illuminating discussions on the topology optimization method, and to 
Krister Svanberg for providing us with the \Matlab{} code for the MMA 
algorithm.
This work is partly supported by The Danish Technical Research Council, 
Grant No.~26-03-0037.

\appendix
\section*{Appendix}
{\footnotesize\begin{verbatim}
% FEMLAB CODE FOR THE 4-TERMINAL DEVICE EXAMPLE OF SEC. 4.2
clear fem femadj
% DEFINE REYNOLDS NUMBER, DARCY NUMBER, LENGTH OF DESIGN DOMAIN, AND VOLUME FRACTION
Re = 50;
Da = 1e-4;
L0 = 3.0;
beta = 0.4;
% DEFINE GEOMETRY, MESH, AND SUBDOMAIN/BOUNDARY GROUPS [SEE FIG. 6]
fem.geom = rect2(0,L0,0,5) + rect2(-2,0,1,2) + rect2(-2,0,3,4) + rect2(L0,L0+2,1,2) ...
   + rect2(L0,L0+2,3,4);
fem.mesh = meshinit(fem,'Hmaxsub',[3 0.125]);
% subdomain groups  1:design domain   2:inlet/outlet leads
fem.equ.ind = {[3] [1 2 4 5]};
% boundary groups  1:walls   2:inlets   3:outlets   4:interior
fem.bnd.ind = {[2:3 5:8 10 12:14 16:18 20:22] [4 23] [1 24] [9 11 15 19]};
% DEFINE SPACE COORDINATES, DEPENDENT VARIABLES, AND SHAPE FUNCTIONS
fem.sdim = {'x' 'y'};
fem.dim = {'u' 'v' 'p' 'gamma'};
fem.shape = [2 2 1 1];
% DEFINE CONSTANTS
fem.const.rho = 1;
fem.const.eta = 1;
fem.const.umax = Re;
fem.const.alphamin = 0;
fem.const.alphamax = 1/Da;
fem.const.q = 0.1;
Phi0 = 96*fem.const.eta*(L0+4)*fem.const.umax^2/9;
% DEFINE EXPRESSIONS ON SUBDOMAIN AND BOUNDARY GROUPS
fem.equ.expr = {'A' 'eta*(2*ux*ux+2*vy*vy+(uy+vx)*(uy+vx))+alpha*(u*u+v*v)' ...
   'alpha' {'alphamin+(alphamax-alphamin)*q*(1-gamma)/(q+gamma)' '0'}};
fem.bnd.expr = {'B' '0'};

% DEFINE GOVERNING EQUATIONS AND INITIAL CONDITIONS [SEE EQS. (8) AND (9)]
fem.form = 'general';
fem.equ.shape = {[1:4] [1:3]};          % only define gamma on subdomain group 1
fem.equ.ga = {{{'-p+2*eta*ux' 'eta*(uy+vx)'} {'eta*(uy+vx)' '-p+2*eta*vy'} {0 0} {0 0}}};
fem.equ.f = {{'rho*(u*ux+v*uy)+alpha*u' 'rho*(u*vx+v*vy)+alpha*v' 'ux+vy' 1}};
fem.equ.init = {{0 0 0 beta}};
% DEFINE BOUNDARY CONDITIONS
fem.bnd.shape = {[1:3]};                % do not define gamma on any boundaries
fem.bnd.r = {{'u' 'v' 0 0} ...          % walls:     no-slip
   {'u*nx+4*umax*s*(1-s)' 'v' 0 0} ...  % inlets:    parabolic profile
   {0 'v' 0 0} ...                      % outlets:   normal flow
   {0 0 0 0}};                          % interior:  nothing
fem.bnd.g = {{0 0 0 0}};                % zero prescribed external forces everywhere
% PERFORM LINEARIZATION, DEGREE-OF-FREEDOM ASSIGNMENT, AND ASSEMBLE INITIAL CONDITION
fem = femdiff(fem);
fem.xmesh = meshextend(fem);
fem.sol = asseminit(fem);

% DEFINE STRUCTURE FOR COMPUTING RIGHT-HAND-SIDE IN ADJOINT PROBLEM [SEE EQ. (29)]
femadj = fem;
femadj.equ.ga = {{{'diff(A,ux)' 'diff(A,uy)'} {'diff(A,vx)' 'diff(A,vy)'} ...
   {'diff(A,px)' 'diff(A,py)'} {'diff(A,gammax)' 'diff(A,gammay)'}}};
femadj.equ.f = {{'diff(A,u)' 'diff(A,v)' 'diff(A,p)' 'diff(A,gamma)'}};
femadj.bnd.g = {{'diff(B,u)' 'diff(B,v)' 'diff(B,p)' 'diff(B,gamma)'}};
femadj.xmesh = meshextend(femadj);
% GET INDICES OF DESIGN VARIABLE IN THE GLOBAL SOLUTION VECTOR (fem.sol.u)
i4 = find(asseminit(fem,'Init',{'gamma' 1},'Out','U'));
% COMPUTE VOLUME BELOW DESIGN VARIABLE BASIS FUNCTIONS
L = assemble(fem,'Out',{'L'});
Vgamma = L(i4);
Vdomain = sum(Vgamma);
% GET INDICES OF VELOCITY-PRESSURE VARIABLES
i123 = find(asseminit(fem,'Init',{'u' 1 'v' 1 'p' 1},'Out','U'));

% DEFINE VARIABLES AND PARAMETERS FOR MMA OPTIMIZATION ALGORITHM [SEE REF. [11,12,13]]
a0 = 1;
a = 0;
c = 20;
d = 0;
xmin = 0;
xmax = 1;
xold = fem.sol.u(i4);
xolder = xold;
low = 0;
upp = 1;

% DESIGN LOOP FOR THE ACTUAL TOPOLOGY OPTIMIZATION
for iter = 1:100
   % SOLVE NAVIER-STOKES FLOW PROBLEM TO UPDATE VELOCITY AND PRESSURE
   fem.sol = femnlin(fem,'Solcomp',{'u' 'v' 'p'},'U',fem.sol.u);
   % SOLVE ADJOINT PROBLEM FOR LAGRANGE MULTIPLIERS
   [K N] = assemble(fem,'Out',{'K' 'N'},'U',fem.sol.u);
   [L M] = assemble(femadj,'Out',{'L' 'M'},'U',fem.sol.u);
   femadj.sol = femlin('In',{'K' K(i123,i123)' 'L' L(i123) 'M' zeros(size(M)) 'N' N(:,i123)});
   % SENSITIVITY ANALYSIS
   gamma = fem.sol.u(i4);
   Phi = postint(fem,'A','Edim',2) + postint(fem,'B','Edim',1);
   dPhidgamma = L(i4) - K(i123,i4)'*femadj.sol.u;
   % PERFORM MMA STEP TO UPDATE DESIGN FIELD
   x = gamma;
   f = Phi/Phi0;                g = gamma'*Vgamma/Vdomain - beta;
   dfdx = dPhidgamma/Phi0;      dgdx = Vgamma'/Vdomain;
   d2fdx2 = zeros(size(gamma)); d2gdx2 = zeros(size(gamma'));
   [xnew,y,z,lambda,ksi,eta,mu,zeta,s,low,upp] = mmasub(1,length(gamma),iter, ...
      x,xmin,xmax,xold,xolder,f,dfdx,d2fdx2,g,dgdx,d2gdx2,low,upp,a0,a,c,d);
   xolder = xold; xold = x; gamma = xnew;
   % TEST CONVERGENCE
   if iter >= 100 | max(abs(gamma-xold)) < 0.01
      break
   end
   % UPDATE DESIGN VARIABLE
   u0 = fem.sol.u; u0(i4) = gamma;
   fem.sol = femsol(u0);
   % DISPLAY RESULTS FOR EACH ITERATION STEP
   disp(sprintf('Iter.:%3d   Obj.: %8.4f   Vol.: %6.3f   Change: %6.3f', ...
      iter,f,xold'*Vgamma,max(abs(xnew-xold))))
   postplot(fem,'arrowdata',{'u' 'v'},'tridata','gamma','trimap','gray')
   axis equal; shg; pause(0.1)
end
\end{verbatim}}

\end{document}